\documentstyle[12pt]{article}
\newcommand {\e} {\mbox{\rm e}}

\newcounter{eq}
\newcounter{sc}

\def\overleftrightarrow#1{\vbox{\ialign{##\crcr
 $\leftrightarrow$\crcr\noalign{\kern-1pt\nointerlineskip}
 $\hfil\displaystyle{#1}\hfil$\crcr}}}

\newlength{\minitwocolumn}
\begin{document}

%%%%%%%%%%%%%%%%%%%%%%%%%%%%%%%%%%%%%%%%%%%%%%%%%%%%%%%%%%%%%%%%%%
%%%%%%%%%%%%%%%%%%%%%%%% Title %%%%%%%%%%%%%%%%%%%%%%%%%%%%%%%%%%%
%%%%%%%%%%%%%%%%%%%%%%%%%%%%%%%%%%%%%%%%%%%%%%%%%%%%%%%%%%%%%%%%%%
\begin{flushright}
DPUR/TH/48\\
March, 2016\\
\end{flushright}
\vspace{20pt}

%\magnification=\magstep1
\pagestyle{empty}
\baselineskip15pt
%\font\cmssB=cmss17
%\font\cmssS=cmss10

\begin{center}
{\large\bf Topological Induced Gravity
\vskip 1mm }

\vspace{20mm}
Ichiro Oda \footnote{E-mail address:\ ioda@phys.u-ryukyu.ac.jp}

\vspace{5mm}
           Department of Physics, Faculty of Science, University of the 
           Ryukyus,\\
           Nishihara, Okinawa 903-0213, Japan.\\

\end{center}

%\maketitle

\vspace{5mm}
\begin{abstract}
We propose a topological model of induced gravity (pregeometry) where both Newton's coupling constant and the cosmological 
constant appear as integration constants in solving field equations. The matter sector of a scalar field is also
considered, and by solving field equations it is shown that various types of cosmological solutions in the FRW universe 
can be obtained.  A detailed analysis is given of the meaning of the BRST transformations, which make the induced gravity
be a topological field theory, by means of the canonical quantization analysis, and the physical reason why such BRST 
transformations are needed in the present formalism is clarified. Finally, we propose a dynamical mechanism for fixing
the Lagrange multiplier fields by following the Higgs mechanism. The present study clearly indicates that the induced gravity 
can be constructed at the classical level without recourse to quantum fluctuations of matter and suggests an interesting 
relationship between the induced gravity and the topological quantum field theory.     
\end{abstract}

\newpage
\pagestyle{plain}
\pagenumbering{arabic}
%\setcounter{page}{1}

%%%%%%%%%%%%%%%%%%%%%%%%%%%%%%%%%%%%%%%%%%%%%%%%%%%%%%%%%%%%%%%%%%
%%%%%%%%%%%%%%%%%%%%%%%% Article %%%%%%%%%%%%%%%%%%%%%%%%%%%%%%%%%
%%%%%%%%%%%%%%%%%%%%%%%%%%%%%%%%%%%%%%%%%%%%%%%%%%%%%%%%%%%%%%%%%%

\rm
%%%%%%%%%%%%%%%%%%%%%%%%%%%%%%%%%%%%%%%%%%%%%%%%%%%%%%%%%%%%%%%%%%%%%
%%%%%%%%%%%%%%%%%%%%%%%%%%%%%%   SEC  1    %%%%%%%%%%%%%%%%%%%%%%%%%%
%%%%%%%%%%%%%%%%%%%%%%%%%%%%%%%%%%%%%%%%%%%%%%%%%%%%%%%%%%%%%%%%%%%%%
\section{Introduction}

In the standard formulation of general gravity, the gravitational interaction is
described by writing down the matter action $S_m$ in a generally covariant form,
and then adding to it the Einstein-Hilbert action plus the cosmological constant
%**   GR action  %%%%%%%%%%%%%%%%%%%%%%%%%%%%%%%%%%%%%%%%%%%%%%%%%%%%%%%%%
\begin{eqnarray}
S_{GR} = \frac{1}{16 \pi G} \int d^4 x  \sqrt{-g} \left( R - 2 \Lambda \right),
\label{GR action}
\end{eqnarray}
%%%%%%%%%%%%%%%%%%%%%%%%%%%%%%%%%%%%%%%%%%%%%%%%%%%%%%%%%%%%%%%%%%%
where $G$ is the Newton coupling constant and $\Lambda$ is the cosmological constant.
Taking variation with respect to the metric tensor $g_{\mu\nu}$ provides us with the
well-known Einstein equations with the cosmological constant
%**   Einstein eq  %%%%%%%%%%%%%%%%%%%%%%%%%%%%%%%%%%%%%%%%%%%%%%%%%%%%%%%%%
\begin{eqnarray}
R_{\mu\nu} - \frac{1}{2} g_{\mu\nu} R + \Lambda g_{\mu\nu} = 8 \pi G T_{\mu\nu},
\label{Einstein eq}
\end{eqnarray}
%%%%%%%%%%%%%%%%%%%%%%%%%%%%%%%%%%%%%%%%%%%%%%%%%%%%%%%%%%%%%%%%%%%
where $T_{\mu\nu}$ is the energy-momentum tensor of matter which is defined as $T_{\mu\nu}
= - \frac{2}{\sqrt{-g}} \frac{\delta S_m}{\delta g^{\mu\nu}}$.  
Although the Einstein equations are known to account very well for various 
gravitational phenomena at the long distance range, the Einstein-Hilbert action (\ref{GR action}) 
yields a non-renormalizable quantum field theory, which has made it difficult to construct quantum
gravity on the basis of the the Einstein-Hilbert action (\ref{GR action}) for a long time.
\footnote{Quantum gravity is considered as a quantum field theory of gravity holding till the
Planck mass scale. In this sense, quantum gravity possesses the proper mass scale in the theory, 
and it is believed that beyond the Planck mass scale, the concept of space-time in itself becomes meaningless.
In such a situation, it is not obvious for at least the present author whether quantum gravity should be described 
in terms of a renormalizable theory or not since the concept of the renormalizability in quantum field theories 
does not suppose the existence of the upper limit value of energy like the Planck mass scale.}

An entirely different approach to quantum gravity, which is nowdays called induced gravity or
pregeometry, has stemmed from a remarkable idea by Sakharov in 1968 \cite{Sakharov}. His basic idea is 
that gravity is not fundamental but might be induced by quantum fluctuations of matter fields. 
In particular, the Newton coupling constant could be generated at the one-loop level although it is vanishing
at the tree level. 

In the framework of the induced gravity, the existence of the Lorentzian manifold
is assumed $\it{a \ priori}$, but the dynamics of the geometry is not assumed but
determined by radiative corrections of matter. Thus, the geometry is regarded as
a classical and fixed background so that the metric tensor field $g_{\mu\nu}$ 
is not quantized unlike matter fields. Being expressed by using
an equation, integrating over matter fluctuations at the one-loop level turns out to produce an effective action
%**   Effective action  %%%%%%%%%%%%%%%%%%%%%%%%%%%%%%%%%%%%%%%%%%%%%%%%%%%%%%%%%
\begin{eqnarray}
S_{IG} = \frac{1}{16 \pi G} \int d^4 x  \sqrt{-g} \left( - 2 \Lambda + R 
+ \cdots \right),
\label{Effective action}
\end{eqnarray}
%%%%%%%%%%%%%%%%%%%%%%%%%%%%%%%%%%%%%%%%%%%%%%%%%%%%%%%%%%%%%%%%%%%
where dots denote the higher-derivative terms such as $R^2$. Even if there are still some difficulties and
subtleties in the induced gravity, the basic idea is extremely of interest in the sense
that quantum gravity is now believed to be somehow an emergent phenomenon like the induced 
gravity \cite{Seiberg}.
   
There are a number of versions of the induced gravity thus far \cite{Akama}-\cite{Frolov}, \footnote{See
a recent review on induced gravity \cite{Visser}.}
but a common feature of all the induced gravity is to derive general relativity or its generalizations 
via quantum corrections of matter (or gauge field).
However, there might be a logical possibility in such a way that without recourse to quantum fluctuations
of matter, gravity could be generated by solving field equations at the classical level. 
Here we have the pregeometrical approach of string field theory \cite{Hata, Horowitz} in mind.
In this approach of string theory, the dynamics of the geometry or the kinetic term involving 
a background metric is created by starting with only a purely cubic action and then picking up 
a suitable classical solution where quantum corrections play no role. 
One of motivations behind the study at hand is to construct such an induced gravity within the framework 
of a local field theory.     
  
Actually, in recent works \cite{Oda3, Oda4}, motivated by ideas of unimodular gravity 
\cite{Einstein}-\cite{Carballo-Rubio}, 
we have constructed a new model of the induced gravity where the Newton coupling constant 
and the cosmological constant simultaneously appear as integration constants in solving field equations. 
Even if these integration constants take any values and are not determined uniquely from a fundamental
principle, choosing appropriate values automatically lead to the Einstein equations with the cosmological 
constant. In this paper, we will generalize this formulation of the induced gravity by including a matter field
and try to find a cosmological solution describing an acceleratingly expanding universe like the present universe.
Moreover, we will propose a dynamical mechanism for fixing the Lagrange multiplier fields to physically
plausible values via spontaneous symmetry breakdown.

The structure of this article is the following: In Section 2, we generalize the previous model \cite{Oda3, Oda4}
of the induced gravity by including the matter action of a scalar field. Then adding a ghost sector, we construct 
a topological field theory. In Section 3, a detailed analysis is given of the meaning of the BRST transfomations 
which were constructed in Section 2. 
In Section 4, we examine the cosmological implications by finding classical solutions to the field
equations stemming from the topological model.
In Section 5, according to the Higgs mechanism, we propose a dynamical mechanism for fixing the
Lagrange multiplier fields to be suitable values.
We conclude in Section 6.

%%%%%%%%%%%%%%%%%%%%%%%%%%%%%%%%%%%%%%%%%%%%%%%%%%%%%%%%%%%%%%%%%%%%%
%%%%%%%%%%%%%%%%%%%%%%%%%%%%%%   SEC  2    %%%%%%%%%%%%%%%%%%%%%%%%%%
%%%%%%%%%%%%%%%%%%%%%%%%%%%%%%%%%%%%%%%%%%%%%%%%%%%%%%%%%%%%%%%%%%%%%
\section{New type of induced gravity}

We start by presenting an action of a new type of induced gravity model: 
\footnote{We follow notation and conventions by Misner et al.'s textbook \cite{MTW}, 
for instance, the flat Minkowski metric $\eta_{\mu\nu} = diag(-, +, +, +)$, the Riemann curvature tensor 
$R^\mu \ _{\nu\alpha\beta} = \partial_\alpha \Gamma^\mu_{\nu\beta} - \partial_\beta \Gamma^\mu_{\nu\alpha} 
+ \Gamma^\mu_{\sigma\alpha} \Gamma^\sigma_{\nu\beta} - \Gamma^\mu_{\sigma\beta} \Gamma^\sigma_{\nu\alpha}$, 
and the Ricci tensor $R_{\mu\nu} = R^\alpha \ _{\mu\alpha\nu}$. The reduced Planck mass is defined as 
$M_p = \sqrt{\frac{c \hbar}{8 \pi G}} = 2.4 \times 10^{18} GeV$ where $G$ is the Newton coupling constant.
Through this article, we adopt the reduced Planck units where we set $c = \hbar = M_p = 1$. 
In this units, all quantities become dimensionless. The Newton coupling constant $G$ is sometimes restored 
when it is necessary.}
%**   IG-action 1  %%%%%%%%%%%%%%%%%%%%%%%%%%%%%%%%%%%%%%%%%%%%%%%%%%%%%%%%%
\begin{eqnarray}
S = \int d^4 x  \sqrt{-g} \left\{ \gamma \left( R + \nabla_\mu \tau^\mu \right)
+ \lambda \left( 1 + \nabla_\mu \omega^\mu \right) 
+ \beta \left[ - \frac{1}{2} g^{\mu\nu} \nabla_\mu \phi \nabla_\nu \phi 
- V(\phi) + \nabla_\mu \rho^\mu \right] \right\},
\label{IG-action 1}
\end{eqnarray}
%%%%%%%%%%%%%%%%%%%%%%%%%%%%%%%%%%%%%%%%%%%%%%%%%%%%%%%%%%%%%%%%%%%
where $\gamma(x)$, $\lambda(x)$ and $\beta(x)$ are the Lagrange multiplier fields enforcing the following
constraints, respectively: 
%**   IG-constraints 1  %%%%%%%%%%%%%%%%%%%%%%%%%%%%%%%%%%%%%%%%%%%%%%%%%%%%%%%%%
\begin{eqnarray}
R = - \nabla_\mu \tau^\mu, \quad 1 = - \nabla_\mu \omega^\mu, 
\quad \frac{1}{2} g^{\mu\nu} \nabla_\mu \phi \nabla_\nu \phi 
+ V(\phi) = \nabla_\mu \rho^\mu.
\label{IG-constraints 1}
\end{eqnarray}
%%%%%%%%%%%%%%%%%%%%%%%%%%%%%%%%%%%%%%%%%%%%%%%%%%%%%%%%%%%%%%%%%%%
Here $\tau^\mu$, $\omega^\mu$ and $\rho^\mu$ are vector fields, which will be fixed in searching for
classical solutions. The first term in the action (\ref{IG-action 1}) was introduced to make the curvature density 
constraint $\sqrt{-g} R = 1$ be invariant under not the transverse diffeomorphisms (TDiff) but the full 
diffeomorphisms \cite{Oda1}. The second term was also introduced to keep the unimodular constraint $\sqrt{-g} = 1$
be invariant under diffeomorphisms \cite{Henneaux}. The final term is a new term which introduces the matter action
of a scalar field $\phi$ with a unspecified potential $V(\phi)$.

We are now willing to show that the action (\ref{IG-action 1}) induces the Einstein-Hilbert action
with the cosmological constant as well as the matter action of the scalar field by solving its field equations as follows: 
First, taking variation with respect to the vector fields reads
%**   tau-omega-rho 1  %%%%%%%%%%%%%%%%%%%%%%%%%%%%%%%%%%%%%%%%%%%%%%%%%%%%%%%%%
\begin{eqnarray}
\frac{\delta S}{\delta \tau^\mu} &=& - \sqrt{-g} \nabla_\mu \gamma = 0, \nonumber\\
\frac{\delta S}{\delta \omega^\mu} &=& - \sqrt{-g} \nabla_\mu \lambda = 0, \nonumber\\
\frac{\delta S}{\delta \rho^\mu} &=& - \sqrt{-g} \nabla_\mu \beta = 0,
\label{tau-omega-rho 1}
\end{eqnarray}
%%%%%%%%%%%%%%%%%%%%%%%%%%%%%%%%%%%%%%%%%%%%%%%%%%%%%%%%%%%%%%%%%%%
from which the classical solution is given by
%**   tau-omega-rho 2  %%%%%%%%%%%%%%%%%%%%%%%%%%%%%%%%%%%%%%%%%%%%%%%%%%%%%%%%%
\begin{eqnarray}
 \gamma(x) = \bar \gamma, \quad \lambda(x) = \bar \lambda, \quad \beta(x) = \bar \beta,
\label{tau-omega-rho 2}
\end{eqnarray}
%%%%%%%%%%%%%%%%%%%%%%%%%%%%%%%%%%%%%%%%%%%%%%%%%%%%%%%%%%%%%%%%%%%
where $\bar \gamma$, $\bar \lambda$ and $\bar \beta$ are certain constants. 
Then, provided that we set
%**   tau-omega-rho 3  %%%%%%%%%%%%%%%%%%%%%%%%%%%%%%%%%%%%%%%%%%%%%%%%%%%%%%%%%
\begin{eqnarray}
\bar \gamma = \frac{1}{16 \pi G}, \quad \bar \lambda = - \frac{2 \Lambda}{16 \pi G},
\quad \bar \beta = 1,
\label{tau-omega-rho 3}
\end{eqnarray}
%%%%%%%%%%%%%%%%%%%%%%%%%%%%%%%%%%%%%%%%%%%%%%%%%%%%%%%%%%%%%%%%%%%
we obtain
%**   GR+matter action  %%%%%%%%%%%%%%%%%%%%%%%%%%%%%%%%%%%%%%%%%%%%%%%%%%%%%%%%%
\begin{eqnarray}
S = \int d^4 x  \sqrt{-g} \left[ \frac{1}{16 \pi G} \left( R - 2 \Lambda \right)
- \frac{1}{2} g^{\mu\nu} \nabla_\mu \phi \nabla_\nu \phi - V(\phi) \right],
\label{GR+matter action}
\end{eqnarray}
%%%%%%%%%%%%%%%%%%%%%%%%%%%%%%%%%%%%%%%%%%%%%%%%%%%%%%%%%%%%%%%%%%%
which is equivalent to the Einstein-Hilbert action with the cosmological constant $\Lambda$
plus the scalar matter action. In this way, we can reach the action (\ref{GR+matter action}) 
of general relativity by starting with the action (\ref{IG-action 1}) of the induced gravity 
simply by solving field equations at the classical level. 

A peculiar feature of this new induced gravity is that the Newton constant ($G$)
and the cosmological constant ($\Lambda$) appear as integration constants which are 
not related to any parameters in the original induced gravity action. Let us recall 
that in unimodular gravity \cite{Einstein}-\cite{Carballo-Rubio}, the similar phenomenon
emerges for the cosmological constant and is expected to play an important role in
understanding the well-known cosmological constant problem.

Next, we wish to show that the induced gravity action (\ref{IG-action 1}) can be deformed into 
a topological quantum field theory by adding ghosts and antighosts.

First of all, let us consider three types of the BRST transformations, which are given by
%**   BRST  %%%%%%%%%%%%%%%%%%%%%%%%%%%%%%%%%%%%%%%%%%%%%%%%%%%%%%%%%
\begin{eqnarray}
\delta_B^{(1)} \gamma = \delta_B^{(1)} c^{(1)} = 0, \quad
\delta_B^{(1)} \tau^\mu = \nabla^\mu c^{(1)}, \quad \delta_B^{(1)} b^{(1)} = \gamma, \nonumber\\
\delta_B^{(2)} \lambda = \delta_B^{(2)} c^{(2)} = 0, \quad
\delta_B^{(2)} \omega^\mu = \nabla^\mu c^{(2)}, \quad \delta_B^{(2)} b^{(2)} = \lambda, \nonumber\\
\delta_B^{(3)} \beta = \delta_B^{(3)} c^{(3)} = 0, \quad
\delta_B^{(3)} \rho^\mu = \nabla^\mu c^{(3)}, \quad \delta_B^{(3)} b^{(3)} = \beta,
\label{BRST}
\end{eqnarray}
%%%%%%%%%%%%%%%%%%%%%%%%%%%%%%%%%%%%%%%%%%%%%%%%%%%%%%%%%%%%%%%%%%%
where $c^{(i)}$ and $b^{(i)} (i = 1, 2, 3)$ are respectively ghosts with the ghost number $+1$ and antighosts with 
the ghost number $-1$. One peculiar feature of the BRST transformations (\ref{BRST}) is that the Lagrange mutiplier 
fields $\gamma, \lambda, \beta$ are identified with the Nakanishi-Lautrup auxiliary fields. 
Note that three kinds of the BRST transformations are nilpotent, $\left(\delta_B^{(i)} \right)^2 = 0$ and anticommute 
to each other, $\left\{ \delta_B^{(i)}, {\delta}_B^{(j)} \right\} = 0 \ (i \neq j)$, 
so we can define the physical state, $|phys \rangle$ by the physical state conditions, $Q_B^{(i)} |phys \rangle = 0$ 
where $Q_B^{(i)}$ are the BRST charges \cite{Kugo}.   

Next, let us add the kinetic terms for the ghosts to the action (\ref{IG-action 1}). As a result, the total
action $S_T$ is of form
%**   Total Action   %%%%%%%%%%%%%%%%%%%%%%%%%%%%%%%%%%%%%%%%%%%%%%%%%%%%%%%%%
\begin{eqnarray}
S_T &=& S + \int d^4 x  \sqrt{-g} \sum_{i=1}^3 \nabla_\mu b^{(i)} \nabla^\mu c^{(i)}    \nonumber\\
&=& \int d^4 x  \sqrt{-g} \left\{ \gamma \left( R + \nabla_\mu \tau^\mu \right)
+ \lambda \left( 1 + \nabla_\mu \omega^\mu \right) 
+ \beta \Bigl[ - \frac{1}{2} g^{\mu\nu} \nabla_\mu \phi \nabla_\nu \phi 
- V(\phi) + \nabla_\mu \rho^\mu \Bigr] \right\}
\nonumber\\
&+& \int d^4 x  \sqrt{-g} \sum_{i=1}^3 \nabla_\mu b^{(i)} \nabla^\mu c^{(i)}  
\nonumber\\
&=& \int d^4 x  \sqrt{-g} \Biggl\{ \delta_B^{(1)} \Bigl[ b^{(1)} \left( R +
\nabla_\mu \tau^\mu \right) \Bigr] 
+ \delta_B^{(2)} \left[ b^{(2)} \left( 1 + \nabla_\mu \omega^\mu \right) \right] 
\nonumber\\
&+&  \delta_B^{(3)} \Bigl[  b^{(3)} \Bigl( - \frac{1}{2} g^{\mu\nu} 
\nabla_\mu \phi \nabla_\nu \phi - V(\phi) 
+ \nabla_\mu \rho^\mu \Bigr) \Bigr] \Biggr\}.
\label{Total Action}
\end{eqnarray}
%%%%%%%%%%%%%%%%%%%%%%%%%%%%%%%%%%%%%%%%%%%%%%%%%%%%%%%%%%%%%%%%%%%
The last expression clearly implies that $S_T$ is invariant under three BRST transformations (\ref{BRST})
and it is an action of the topological quantum field theory whose moduli space is defined by three equations 
(\ref{IG-constraints 1}).

About a quarter of a century ago, starting with a trivially zero classical action, we have established a model 
of topological pregeomery where quantum fluctuations of matter fields have generated the Einstein-Hilbert 
action with the cosmological constant at the one-loop level via the cutoff which violates the topological 
symmetry \cite{Oda1, Oda2}. 
On the other hand, in this paper, we have started with a non-trivial classical action (\ref{IG-action 1}) 
of the induced gravity, which reduces to the Einstein-Hilbert action with the cosmological constant as well as
an action of a scalar field at the classical level, and then added the kinetic term of the ghosts, thereby 
transforming the induced gravity to a topological quantum field theory. The above two models of the induced gravity 
are obviously different, but the close relationship between the induced gravity and the topological field theory 
seems to shed some light on the further development of the two theories in future.

%%%%%%%%%%%%%%%%%%%%%%%%%%%%%%%%%%%%%%%%%%%%%%%%%%%%%%%%%%%%%%%%%%%%%
%%%%%%%%%%%%%%%%%%%%%%%%%%%%%%   SEC  3    %%%%%%%%%%%%%%%%%%%%%%%%%%
%%%%%%%%%%%%%%%%%%%%%%%%%%%%%%%%%%%%%%%%%%%%%%%%%%%%%%%%%%%%%%%%%%%%%
\section{Meaning of BRST transformations}

In the previous section, we have presented three BRST transformations (\ref{BRST}) which are essential 
in deforming our induced gravity action to a topological quantum field theory. Usually, a topological symmetry 
is equivalent to a local shift symmetry by which the corresponding field can be gauge-fixed to zero at
least locally. On the other hand, our topological symmetry is similar to the gauge symmetry of a vector
field, so it is not clear whether the BRST transformations (\ref{BRST}) correspond to the conventional
topological symmetry. In this section, we will demonstrate that this is indeed the case by performing the canonical
quantization of a toy model, which is similar to our induced gravity action, in a flat Minkowski space-time.

Let us focus on a toy model, which captures the essential features of the BRST transformation of the induced 
gravity (\ref{IG-action 1}), in a flat Minkowski space-time in order to avoid complications associated with
a curved space-time. The generalization to the curved space-time is straightforward and explained later.
We begin with the following Lagrange density: 
%**   Toy model 1   %%%%%%%%%%%%%%%%%%%%%%%%%%%%%%%%%%%%%%%%%%%%%%%%%%%%%%%%%
\begin{eqnarray}
{\cal{L}} = \gamma \left[ \Psi(\varphi) + \partial_\mu \tau^\mu \right] + \partial_\mu b \partial^\mu c,
\label{Toy model 1}
\end{eqnarray}
%%%%%%%%%%%%%%%%%%%%%%%%%%%%%%%%%%%%%%%%%%%%%%%%%%%%%%%%%%%%%%%%%%%
where $\gamma$ and $\tau^\mu$ are respectively a scalar and a vector field, $\Psi$ is a generic function 
of a general field $\varphi$, and then $b$ and $c$ are respectively an antighost and a ghost.

This Lagrangian density (\ref{Toy model 1}) is invariant up to a surface term under the following 
nilpotent BRST transformation: 
%**   Simple BRST  %%%%%%%%%%%%%%%%%%%%%%%%%%%%%%%%%%%%%%%%%%%%%%%%%%%%%%%%%
\begin{eqnarray}
\delta_B \gamma = \delta_B \varphi = \delta_B c = 0, \quad 
\delta_B \tau^\mu = \partial^\mu c, \quad \delta_B b = \gamma.
\label{Simple BRST}
\end{eqnarray}
%%%%%%%%%%%%%%%%%%%%%%%%%%%%%%%%%%%%%%%%%%%%%%%%%%%%%%%%%%%%%%%%%%%
It is easy to rewrite (\ref{Toy model 1}) to a BRST-exact form
%**   BRST-exact   %%%%%%%%%%%%%%%%%%%%%%%%%%%%%%%%%%%%%%%%%%%%%%%%%%%%%%%%%
\begin{eqnarray}
{\cal{L}} = \delta_B \left\{ b \left[ \Psi(\varphi) + \partial_\mu \tau^\mu \right] \right\}.
\label{BRST-exact}
\end{eqnarray}
%%%%%%%%%%%%%%%%%%%%%%%%%%%%%%%%%%%%%%%%%%%%%%%%%%%%%%%%%%%%%%%%%%%
In what follows, we will ignore the ghost sector since this sector is irrelevant to the following argument.
Moreover, we assume that the general field $\varphi$ is a fixed background field so that we do not regard it 
as the dynamical variable. 

We will make use of the canonical quantization procedure to clarify a hidden gauge symmetry in the Lagrangian
density (\ref{Toy model 1}). First, let us introduce the canonical conjugate momenta $\pi_\mu$ corresponding
to the canonical variables $\tau^\mu$ \footnote{Here we regard $\gamma$ as the canonical conjugate momentum
for $\tau^0$ rather than the canonical variable. It is possible to derive the same result by using 
the Dirac bracket \cite{Dirac} even if we deal with $\gamma$ as the canonical variable.}
%**   Canonical momenta   %%%%%%%%%%%%%%%%%%%%%%%%%%%%%%%%%%%%%%%%%%%%%%%%%%%%%%%%%
\begin{eqnarray}
\pi_0 = \frac{\partial {\cal{L}}}{\partial \dot{\tau}^0} = \gamma, \quad
\pi_i = \frac{\partial {\cal{L}}}{\partial \dot{\tau}^i} = 0,
\label{Canonical momenta}
\end{eqnarray}
%%%%%%%%%%%%%%%%%%%%%%%%%%%%%%%%%%%%%%%%%%%%%%%%%%%%%%%%%%%%%%%%%%%
where the overdot means the differentiation with respect to $x^0 = t$, i.e., $\dot{\tau}^0 = \partial_0 \tau^0$ 
and the index $i$ denotes the spacial 
components $i = 1, 2, 3$. The latter equations imply that we have primary constraints
%**   Primary constraint   %%%%%%%%%%%%%%%%%%%%%%%%%%%%%%%%%%%%%%%%%%%%%%%%%%%%%%%%%
\begin{eqnarray}
\pi_i \approx 0.
\label{Primary constraint}
\end{eqnarray}
%%%%%%%%%%%%%%%%%%%%%%%%%%%%%%%%%%%%%%%%%%%%%%%%%%%%%%%%%%%%%%%%%%%
The Hamiltonian density reads
%**   Hamiltonian   %%%%%%%%%%%%%%%%%%%%%%%%%%%%%%%%%%%%%%%%%%%%%%%%%%%%%%%%%
\begin{eqnarray}
{\cal{H}} = - \pi_0 \left[ \Psi(\varphi) + \partial_i \tau^i \right].
\label{Hamiltonian}
\end{eqnarray}
%%%%%%%%%%%%%%%%%%%%%%%%%%%%%%%%%%%%%%%%%%%%%%%%%%%%%%%%%%%%%%%%%%%

At this stage, we set up the equal-time commutation relations:
%**   CCR   %%%%%%%%%%%%%%%%%%%%%%%%%%%%%%%%%%%%%%%%%%%%%%%%%%%%%%%%%
\begin{eqnarray}
\left[ \tau^\mu (t, \vec{x}), \pi_\nu (t, \vec{y}) \right] = i \delta^\mu_\nu \delta^3 (x - y). 
\label{CCR}
\end{eqnarray}
%%%%%%%%%%%%%%%%%%%%%%%%%%%%%%%%%%%%%%%%%%%%%%%%%%%%%%%%%%%%%%%%%%%
The time evolution of the primary constraints (\ref{Primary constraint}) produces the secondary constraints
%**   Secondary constraint   %%%%%%%%%%%%%%%%%%%%%%%%%%%%%%%%%%%%%%%%%%%%%%%%%%%%%%%%%
\begin{eqnarray}
\partial_i \pi_0 \approx 0.
\label{Secondary constraint}
\end{eqnarray}
%%%%%%%%%%%%%%%%%%%%%%%%%%%%%%%%%%%%%%%%%%%%%%%%%%%%%%%%%%%%%%%%%%%
It is straightforward to show that there is no ternary constraint. Thus, the whole constraint is
given by the sum of the primary constraints (\ref{Primary constraint}) and the secondary constraints 
(\ref{Secondary constraint}), and since these constraints commute with each other under the commutation
relations (\ref{CCR}), 
they constitute the first-class constraints, which generate gauge transformations. Actually, if we define
a generating function of the gauge transformations by 
%**   Generator   %%%%%%%%%%%%%%%%%%%%%%%%%%%%%%%%%%%%%%%%%%%%%%%%%%%%%%%%%
\begin{eqnarray}
\Phi = i \int d^3 x  \left( \epsilon^i \pi_i + \xi^i \partial_i \pi_0 \right),
\label{Generator}
\end{eqnarray}
%%%%%%%%%%%%%%%%%%%%%%%%%%%%%%%%%%%%%%%%%%%%%%%%%%%%%%%%%%%%%%%%%%%
we can derive the gauge transformations associated with the constraints
%**   Fake gauge transf   %%%%%%%%%%%%%%%%%%%%%%%%%%%%%%%%%%%%%%%%%%%%%%%%%%%%%%%%%
\begin{eqnarray}
\left[ \Phi, \tau^0 \right] = - \partial_i \xi^i, \quad
\left[ \Phi, \tau^i \right] = \epsilon^i. 
\label{Fake gauge transf}
\end{eqnarray}
%%%%%%%%%%%%%%%%%%%%%%%%%%%%%%%%%%%%%%%%%%%%%%%%%%%%%%%%%%%%%%%%%%%
However, all the parameters $\xi^i, \epsilon^i$ are not always independent.
The Lorentz invariance suggests us to take $\xi^i = \dot{\epsilon}^i$. Hence, 
the gauge transformations are given by
%**   Real gauge transf   %%%%%%%%%%%%%%%%%%%%%%%%%%%%%%%%%%%%%%%%%%%%%%%%%%%%%%%%%
\begin{eqnarray}
\delta \tau^0 = - \partial_i \xi^i, \quad
\delta \tau^i = \partial_0 \xi^i. 
\label{Real gauge transf}
\end{eqnarray}
%%%%%%%%%%%%%%%%%%%%%%%%%%%%%%%%%%%%%%%%%%%%%%%%%%%%%%%%%%%%%%%%%%%
In fact, it is easy to check that the Lagrangian density (\ref{Toy model 1}) is invariant
under the gauge transformations (\ref{Real gauge transf}).

The above gauge transformations mean that three components in $\tau^\mu$ can be
gauged away and only one component is an independent variable. Then, the Lorentz
invariance again insists that the one independent component should be a scalar field $\theta$
described as $\tau^\mu = \partial^\mu \theta$. Substituting this expression for $\tau^\mu$
into the starting Langrangian density (\ref{Toy model 1}) gives rise to
%**   Toy model 2   %%%%%%%%%%%%%%%%%%%%%%%%%%%%%%%%%%%%%%%%%%%%%%%%%%%%%%%%%
\begin{eqnarray}
{\cal{L}} = \gamma \left[ \Psi(\varphi) + \partial_\mu \partial^\mu \theta \right].
\label{Toy model 2}
\end{eqnarray}
%%%%%%%%%%%%%%%%%%%%%%%%%%%%%%%%%%%%%%%%%%%%%%%%%%%%%%%%%%%%%%%%%%%
Note that we have no longer gauge symmetry in (\ref{Toy model 2}).

After integrating by parts, the Lagrangian (\ref{Toy model 2}) density is recast to the form 
%**   Toy model 3   %%%%%%%%%%%%%%%%%%%%%%%%%%%%%%%%%%%%%%%%%%%%%%%%%%%%%%%%%
\begin{eqnarray}
{\cal{L}} = \gamma \Psi(\varphi) - \partial_\mu \gamma \partial^\mu \theta.
\label{Toy model 3}
\end{eqnarray}
%%%%%%%%%%%%%%%%%%%%%%%%%%%%%%%%%%%%%%%%%%%%%%%%%%%%%%%%%%%%%%%%%%%
The presence of the mixing kinetic term $- \partial_\mu \gamma \partial^\mu \theta$ suggests the 
existence of a ghost. Indeed, when we introduce two scalar fields $\phi_1, \phi_2$ instead of
$\gamma, \theta$ by
%**   gamma-theta   %%%%%%%%%%%%%%%%%%%%%%%%%%%%%%%%%%%%%%%%%%%%%%%%%%%%%%%%%
\begin{eqnarray}
\gamma = \frac{1}{\sqrt{2}} \left( \phi_1 + \phi_2 \right),  \quad
\theta = \frac{1}{\sqrt{2}} \left( \phi_1 - \phi_2 \right),
\label{gamma-theta}
\end{eqnarray}
%%%%%%%%%%%%%%%%%%%%%%%%%%%%%%%%%%%%%%%%%%%%%%%%%%%%%%%%%%%%%%%%%%%
the Lagrangian density reads
%**   Toy model 4   %%%%%%%%%%%%%%%%%%%%%%%%%%%%%%%%%%%%%%%%%%%%%%%%%%%%%%%%%
\begin{eqnarray}
{\cal{L}} =  \frac{1}{\sqrt{2}} \left( \phi_1 + \phi_2 \right) \Psi(\varphi) 
- \frac{1}{2} \left( \partial_\mu \phi_1 \partial^\mu \phi_1
- \partial_\mu \phi_2 \partial^\mu \phi_2 \right).
\label{Toy model 4}
\end{eqnarray}
%%%%%%%%%%%%%%%%%%%%%%%%%%%%%%%%%%%%%%%%%%%%%%%%%%%%%%%%%%%%%%%%%%%
This form of the Lagrangian density clearly shows that $\phi_2$ is a ghost field whereas 
$\phi_1$ is a normal scalar field. 

One reasonable way to remove the ghost $\phi_2$, which violates the unitarity of the theory, from the physical spectrum
is to introduce a topological symmetry like
%**   New symmetry   %%%%%%%%%%%%%%%%%%%%%%%%%%%%%%%%%%%%%%%%%%%%%%%%%%%%%%%%%
\begin{eqnarray}
\delta \theta (x) = \varepsilon (x).
\label{New symmetry}
\end{eqnarray}
%%%%%%%%%%%%%%%%%%%%%%%%%%%%%%%%%%%%%%%%%%%%%%%%%%%%%%%%%%%%%%%%%%%
Then, the corresponding BRST transformation is obtained by replacing the gauge parameter $\varepsilon (x)$ with
a ghost, which we call $c(x)$. Accordingly, the resulting BRST transfomation for 
$\tau^\mu = \partial^\mu \theta$ takes the form
%**   New BRST   %%%%%%%%%%%%%%%%%%%%%%%%%%%%%%%%%%%%%%%%%%%%%%%%%%%%%%%%%
\begin{eqnarray}
\delta_B \tau^\mu = \partial^\mu c.
\label{New BRST}
\end{eqnarray}
%%%%%%%%%%%%%%%%%%%%%%%%%%%%%%%%%%%%%%%%%%%%%%%%%%%%%%%%%%%%%%%%%%%
This is nothing but the BRST transformation which appears in Eq. (\ref{BRST}), in case of a flat 
Minkowski space-time. In this way, we have shown that the BRST transformations (\ref{BRST}) are really those 
corresponding to topological symmetries.

So far, for simplicity we have limited ourselves to the toy model (\ref{Toy model 1}) in a flat
Minkowski space-time. Although an analysis beomes a little intricated owing to complications
associated with the property of the curved space-time, the above result remains valid with a suitable adjustment
even in a curved space-time. For instance, in a curved space-time, in case of the absence of $b-c$ ghosts
the Lagrangian density (\ref{Toy model 1}) becomes 
%**   Curved toy model 1   %%%%%%%%%%%%%%%%%%%%%%%%%%%%%%%%%%%%%%%%%%%%%%%%%%%%%%%%%
\begin{eqnarray}
{\cal{L}} &=& \sqrt{-g} \gamma \left[ \Psi(\varphi) + \nabla_\mu \tau^\mu \right]  \nonumber\\
&=& \sqrt{-g} \gamma \Psi(\varphi) + \gamma \partial_\mu \left( \sqrt{-g} \tau^\mu \right)     \nonumber\\
&=& \sqrt{-g} \gamma \Psi(\varphi) + \gamma \left[ \partial_0 \left( \sqrt{-g} \tau^0 \right)
+ \partial_i \left( \sqrt{-g} \tau^i \right) \right],
\label{Curved toy model 1}
\end{eqnarray}
%%%%%%%%%%%%%%%%%%%%%%%%%%%%%%%%%%%%%%%%%%%%%%%%%%%%%%%%%%%%%%%%%%%
where at the second equality we have used a formula $\nabla_\mu (\sqrt{-g} A^\mu) = \partial_\mu (\sqrt{-g} A^\mu)$,
which holds for an arbitrary vector field $A^\mu$.
From the last expression, it is easy to see that this Lagrangian density in the curved space-time is invariant 
under gauge transformations
%**   Curved gauge transf   %%%%%%%%%%%%%%%%%%%%%%%%%%%%%%%%%%%%%%%%%%%%%%%%%%%%%%%%%
\begin{eqnarray}
\delta \tau^0 = - \frac{1}{\sqrt{-g}} \partial_i \xi^i, \quad
\delta \tau^i = \frac{1}{\sqrt{-g}} \partial_0 \xi^i,
\label{Curved gauge transf}
\end{eqnarray}
%%%%%%%%%%%%%%%%%%%%%%%%%%%%%%%%%%%%%%%%%%%%%%%%%%%%%%%%%%%%%%%%%%%
where $\xi^i$ are three gauge parameters. These gauge transformations are the natural extension of Eq. (\ref{Real gauge transf})
to a curved space-time. Because of these gauge symmetries and the general covariance, we can set 
$\tau^\mu = \nabla^\mu \theta$. Along the same line of argument as in the case of a flat Minkowski space-time, 
it turns out that the resultant BRST transformation agrees with that of (\ref{BRST}).

%%%%%%%%%%%%%%%%%%%%%%%%%%%%%%%%%%%%%%%%%%%%%%%%%%%%%%%%%%%%%%%%%%%%%
%%%%%%%%%%%%%%%%%%%%%%%%%%%%%%   SEC  4    %%%%%%%%%%%%%%%%%%%%%%%%%%
%%%%%%%%%%%%%%%%%%%%%%%%%%%%%%%%%%%%%%%%%%%%%%%%%%%%%%%%%%%%%%%%%%%%%
\section{Cosmological solutions}

In this section, we work with the action (\ref{Total Action}) of a topological quantum field theory to
derive various cosmological solutions as classical solutions in the framework of the Friedmann-Robertson-Walker 
(FRW) universe with spacially flat metric. This derivation suggests not only that the topological action 
(\ref{Total Action}) is equivalent to that of general relativity with matter but also that it can be 
applied to cosmology.

In particular, we will present two types of classical solutions. One of them is the classical solution 
which is a natural generalization found in our previous paper \cite{Oda4} where the Hubble parameter is
a constant.  This type of the solution describes the inflation universe. The other solution exists only when a scalar 
matter is coupled to gravity and is characterized by the fact that the Hubble parameter explicitly depends on the time variable.
In this case, as in general relativity with a scalar matter, the solution describes various cosmological
solutions, which involve, for instance, an acceleratingly expanding universe like the present universe.

Before doing that, let us first notice that using the gauge symmetries (\ref{Curved gauge transf}) which exist
for each vector field $\tau^\mu, \omega^\mu, \rho^\mu$, one can set spacial components of each vector field
to be vanishing:
%**   Spacial vector  %%%%%%%%%%%%%%%%%%%%%%%%%%%%%%%%%%%%%%%%%%%%%%%%%%%%%%%%%
\begin{eqnarray}
\tau_i = \omega_i = \rho_i = 0.
\label{Spacial vector}
\end{eqnarray}
%%%%%%%%%%%%%%%%%%%%%%%%%%%%%%%%%%%%%%%%%%%%%%%%%%%%%%%%%%%%%%%%%%%
Moreover, as in the conventional approach of cosmology, one assumes that a scalar field $\phi$ as well as
the time component of the above vector fields depend on only the time variable $t$:
%**   t-dependence  %%%%%%%%%%%%%%%%%%%%%%%%%%%%%%%%%%%%%%%%%%%%%%%%%%%%%%%%%
\begin{eqnarray}
\phi = \phi(t), \quad \tau_0 = \tau_0(t), \quad \omega_0 = \omega_0(t), \quad
\rho_0 = \rho_0(t).
\label{t-dependence}
\end{eqnarray}
%%%%%%%%%%%%%%%%%%%%%%%%%%%%%%%%%%%%%%%%%%%%%%%%%%%%%%%%%%%%%%%%%%%
Next, having the cosmological application of the model at hand in mind, let us work with the FRW metric 
with spacially flat metric ($k = 0$)
%**   FRW metric  %%%%%%%%%%%%%%%%%%%%%%%%%%%%%%%%%%%%%%%%%%%%%%%%%%%%%%%%%
\begin{eqnarray}
d s^2 = g_{\mu\nu} d x^\mu d x^\nu = - d t^2 + a(t)^2 ( d x^2 + d y^2 + d z^2 ),
\label{FRW metric}
\end{eqnarray}
%%%%%%%%%%%%%%%%%%%%%%%%%%%%%%%%%%%%%%%%%%%%%%%%%%%%%%%%%%%%%%%%%%%
with $a(t)$ being the scale factor. 

We are now ready to turn our attention to finding classical solutions to field equations, which are obtained from 
the topological action (\ref{Total Action}). We first attempt to search for a classical solution such that the Hubble
parameter is a mere constant.
Variation with respect to the vector fields $\tau^\mu, \omega^\mu, \rho^\mu$ yields Eqs. (\ref{tau-omega-rho 1}), 
from which we have the classical solution (\ref{tau-omega-rho 2}).
In what follows, among any values of constants, we will take the specific value (\ref{tau-omega-rho 3})
to have an exact correspondence with general relativity. 
Field equations for the ghosts and the antighosts satisfy the same form of equations
%**   Ghost eq  %%%%%%%%%%%%%%%%%%%%%%%%%%%%%%%%%%%%%%%%%%%%%%%%%%%%%%%%%
\begin{eqnarray}
\nabla_\mu \nabla^\mu c^{(i)} =  \nabla_\mu \nabla^\mu b^{(i)} = 0.
\label{Ghost eq}
\end{eqnarray}
%%%%%%%%%%%%%%%%%%%%%%%%%%%%%%%%%%%%%%%%%%%%%%%%%%%%%%%%%%%%%%%%%%%
Since the ghosts and the antighosts carry non-zero ghost numbers, we simply take $c^{(i)} = b^{(i)} =0$ 
as the classical solution in this paper.

The Einstein equations, which stem from variation with respect to the metric tensor $g^{\mu\nu}$, read
%**   Einstein-eq 1  %%%%%%%%%%%%%%%%%%%%%%%%%%%%%%%%%%%%%%%%%%%%%%%%%%%%%%%%%
\begin{eqnarray}
&{}& 2 \gamma G_{\mu\nu} - 2 \left( \nabla_\mu \nabla_\nu \gamma - g_{\mu\nu} \nabla^2 \gamma \right)
\nonumber\\
&+& 2 \left[ - \tau_{(\mu} \nabla_{\nu)} \gamma - \omega_{(\mu} \nabla_{\nu)} \lambda 
- \frac{1}{2} \beta \nabla_\mu \phi \nabla_\nu \phi - \rho_{(\mu} \nabla_{\nu)} \beta
+ \sum_i \nabla_{(\mu} b^{(i)} \nabla_{\nu)} c^{(i)} \right]
\nonumber\\
&-& g_{\mu\nu} \left[ \lambda - \tau^\rho \nabla_\rho \gamma - \omega^\rho \nabla_\rho \lambda 
- \beta \left( \frac{1}{2} \nabla_\rho \phi \nabla^\rho \phi + V(\phi) \right)
- \rho^\lambda \nabla_\lambda \beta + \sum_i \nabla_\rho b^{(i)} \nabla^\rho c^{(i)} \right]
\nonumber\\
&=& 0,
\label{Einstein-eq 1}
\end{eqnarray}
%%%%%%%%%%%%%%%%%%%%%%%%%%%%%%%%%%%%%%%%%%%%%%%%%%%%%%%%%%%%%%%%%%%
where $G_{\mu\nu} = R_{\mu\nu} - \frac{1}{2} g_{\mu\nu} R$ is the Einstein tensor and the round bracket indicates 
the symmetrization of indices of weight $\frac{1}{2}$ such as $A_{(\mu} B_{\nu)} = \frac{1}{2} \left( A_\mu B_\nu
+ A_\nu B_\mu \right)$. In deriving this equation, it is useful to rewrite the action (\ref{Total Action}) by
integrating by parts.
With the solution (\ref{tau-omega-rho 3}) and the vanishing ghosts and antighosts,  
the Einstein equations (\ref{Einstein-eq 1}) take the standard form
%**   Einstein-eq 2  %%%%%%%%%%%%%%%%%%%%%%%%%%%%%%%%%%%%%%%%%%%%%%%%%%%%%%%%%
\begin{eqnarray}
G_{\mu\nu} + \Lambda g_{\mu\nu} = 8 \pi G T^{(\phi)}_{\mu\nu},
\label{Einstein-eq 2}
\end{eqnarray}
%%%%%%%%%%%%%%%%%%%%%%%%%%%%%%%%%%%%%%%%%%%%%%%%%%%%%%%%%%%%%%%%%%%
where the energy-momentum tensor on the RHS is defined as 
%**   Matter-T  %%%%%%%%%%%%%%%%%%%%%%%%%%%%%%%%%%%%%%%%%%%%%%%%%%%%%%%%%
\begin{eqnarray}
T^{(\phi)}_{\mu\nu} = \nabla_\mu \phi \nabla_\nu \phi - \frac{1}{2} g_{\mu\nu} \nabla_\rho \phi \nabla^\rho \phi 
- g_{\mu\nu} V(\phi).
\label{Matter-T}
\end{eqnarray}
%%%%%%%%%%%%%%%%%%%%%%%%%%%%%%%%%%%%%%%%%%%%%%%%%%%%%%%%%%%%%%%%%%%
Using the Bianchi identity in Eq. (\ref{Einstein-eq 2}), we have the conservation law
%**   Conservation-T  %%%%%%%%%%%%%%%%%%%%%%%%%%%%%%%%%%%%%%%%%%%%%%%%%%%%%%%%%
\begin{eqnarray}
\nabla^\mu T^{(\phi)}_{\mu\nu} = 0.
\label{Conservation-T}
\end{eqnarray}
%%%%%%%%%%%%%%%%%%%%%%%%%%%%%%%%%%%%%%%%%%%%%%%%%%%%%%%%%%%%%%%%%%%
For later convenience, it is useful to introduce two quantities
%**   rho-p  %%%%%%%%%%%%%%%%%%%%%%%%%%%%%%%%%%%%%%%%%%%%%%%%%%%%%%%%%
\begin{eqnarray}
\rho_{\phi} = \frac{1}{2} \dot{\phi}^2 + V(\phi),    \nonumber\\
p_{\phi} = \frac{1}{2} \dot{\phi}^2 - V(\phi).
\label{rho-p}
\end{eqnarray}
%%%%%%%%%%%%%%%%%%%%%%%%%%%%%%%%%%%%%%%%%%%%%%%%%%%%%%%%%%%%%%%%%%%

It turns out that with the metric ansatz (\ref{FRW metric}), the Einstein's equations (\ref{Einstein-eq 2}) give us 
two sets of the independent equations 
%**   Einstein-eq 3  %%%%%%%%%%%%%%%%%%%%%%%%%%%%%%%%%%%%%%%%%%%%%%%%%%%%%%%%%
\begin{eqnarray}
3 H^2 - \Lambda &=& 8 \pi G \rho_{\phi},    \nonumber\\
- \left( 3 H^2 + 2 \dot{H} \right) + \Lambda &=& 8 \pi G p_{\phi}.
\label{Einstein-eq 3}
\end{eqnarray}
%%%%%%%%%%%%%%%%%%%%%%%%%%%%%%%%%%%%%%%%%%%%%%%%%%%%%%%%%%%%%%%%%%%
Adding the two equations, we have
%**   Einstein-eq 4  %%%%%%%%%%%%%%%%%%%%%%%%%%%%%%%%%%%%%%%%%%%%%%%%%%%%%%%%%
\begin{eqnarray}
- 2 \dot{H} = 8 \pi G \dot{\phi}^2.
\label{Einstein-eq 4}
\end{eqnarray}
%%%%%%%%%%%%%%%%%%%%%%%%%%%%%%%%%%%%%%%%%%%%%%%%%%%%%%%%%%%%%%%%%%%
If we wish to obtain a solution where the Hubble parameter $H(t)$ is a constant, this equation
requires us to choose
%**   H-phi 1  %%%%%%%%%%%%%%%%%%%%%%%%%%%%%%%%%%%%%%%%%%%%%%%%%%%%%%%%%
\begin{eqnarray}
H(t) = \bar H, \quad \phi(t) = \bar \phi,
\label{H-phi 1}
\end{eqnarray}
%%%%%%%%%%%%%%%%%%%%%%%%%%%%%%%%%%%%%%%%%%%%%%%%%%%%%%%%%%%%%%%%%%%
where $\bar H$ is a certain constant defined as 
%**   Bar H   %%%%%%%%%%%%%%%%%%%%%%%%%%%%%%%%%%%%%%%%%%%%%%%%%%%%%%%%%
\begin{eqnarray}
\bar H = \frac{1}{3} \sqrt{\Lambda + 8 \pi G V(\bar \phi)},
\label{Bar H}
\end{eqnarray}
%%%%%%%%%%%%%%%%%%%%%%%%%%%%%%%%%%%%%%%%%%%%%%%%%%%%%%%%%%%%%%%%%%%
and $\bar \phi$ is some constant which is commented shortly. Incidentally, the conservation law of the energy-momentum
tensor of the scalar field $\phi$, Eq. (\ref{Conservation-T}), is then identically satisfied since this equation
is not independent of the Einstein equations and can be derived from Eq. (\ref{Einstein-eq 3}).

With the metric ansatz (\ref{FRW metric}), Eq. (\ref{IG-constraints 1}), which is obtained by taking variation of 
the Lagrange multiplier fields, takes the form 
%**   IG-constraints 2  %%%%%%%%%%%%%%%%%%%%%%%%%%%%%%%%%%%%%%%%%%%%%%%%%%%%%%%%%
\begin{eqnarray}
6 \left( 2 H^2 + \dot{H} \right) &=& \dot{\tau_0} + 3 H \tau_0,    \nonumber\\
\dot{\omega}_0 + 3 H \omega_0 &=& 1,    \nonumber\\
p_\phi &=& \dot{\rho}_0 + 3 H \rho_0.
\label{IG-constraints 2}
\end{eqnarray}
%%%%%%%%%%%%%%%%%%%%%%%%%%%%%%%%%%%%%%%%%%%%%%%%%%%%%%%%%%%%%%%%%%%
Since the present solution satisfies Eq. (\ref{H-phi 1}), a simple solution reads
%**   H-phi 2  %%%%%%%%%%%%%%%%%%%%%%%%%%%%%%%%%%%%%%%%%%%%%%%%%%%%%%%%%
\begin{eqnarray}
\tau_0 = 4 \bar H, \quad \omega_0 = \frac{1}{3 \bar H}, 
\quad \rho_0 = - \frac{1}{3 \bar H} V(\bar \phi).
\label{H-phi 2}
\end{eqnarray}
%%%%%%%%%%%%%%%%%%%%%%%%%%%%%%%%%%%%%%%%%%%%%%%%%%%%%%%%%%%%%%%%%%%
Finally, with the metric ansatz (\ref{FRW metric}), variation of $\phi$ leads to
%**   phi-eq 1   %%%%%%%%%%%%%%%%%%%%%%%%%%%%%%%%%%%%%%%%%%%%%%%%%%%%%%%%%
\begin{eqnarray}
\ddot{\phi} + 3 H \dot{\phi} + V^\prime (\phi) = 0.
\label{phi-eq 1}
\end{eqnarray}
%%%%%%%%%%%%%%%%%%%%%%%%%%%%%%%%%%%%%%%%%%%%%%%%%%%%%%%%%%%%%%%%%%%
With $\phi = \bar \phi$, where $\bar \phi$ is a certain constant, this equation provides us with 
a restriction on $\bar \phi$
%**   V(bar phi)   %%%%%%%%%%%%%%%%%%%%%%%%%%%%%%%%%%%%%%%%%%%%%%%%%%%%%%%%%
\begin{eqnarray}
V^\prime (\bar \phi) = 0.
\label{V(bar phi)}
\end{eqnarray}
%%%%%%%%%%%%%%%%%%%%%%%%%%%%%%%%%%%%%%%%%%%%%%%%%%%%%%%%%%%%%%%%%%%
Eq. (\ref{V(bar phi)}) tells us that $\phi = \bar \phi$ is not a mere constant but really an extremum 
of the potential $V(\phi)$ whose existence is assumed in this paper. In this way, we have derived the
classical solution, which is given by (\ref{tau-omega-rho 2}), (\ref{tau-omega-rho 3}), 
(\ref{H-phi 1}), (\ref{Bar H}), and (\ref{H-phi 2}) in addition to
vanishing ghosts and antighosts. With this classical solution, the scale factor takes the form $a(t) 
= a(0) \e^{\bar H t}$, which describes the inflation universe.

Next, let us look for a classical solution having the time-dependent Hubble parameter.
In this case, we also consider a solution given by (\ref{tau-omega-rho 2}), (\ref{tau-omega-rho 3}), 
and vanishing ghosts and antighosts. Furthermore, for simplicity, we set the cosmological constant
to be zero, i.e., $\Lambda = 0$. 

We shall first begin by solving the conservation law Eq. (\ref{Conservation-T}), which is of form
%**   Conservation-T2  %%%%%%%%%%%%%%%%%%%%%%%%%%%%%%%%%%%%%%%%%%%%%%%%%%%%%%%%%
\begin{eqnarray}
\dot{\rho}_\phi + 3 H \rho_\phi \left( 1 + w_\phi \right) = 0,
\label{Conservation-T2}
\end{eqnarray}
%%%%%%%%%%%%%%%%%%%%%%%%%%%%%%%%%%%%%%%%%%%%%%%%%%%%%%%%%%%%%%%%%%%
where we have defined $w_\phi = \frac{p_\phi}{\rho_\phi}$ and we will henceforth assume $w_\phi$ to be a constant
as in the conventional approach of cosmology. Then, it is straightforward to solve the equation (\ref{Conservation-T2})
for $\rho_\phi$ whose expression is 
%**   rho_phi  %%%%%%%%%%%%%%%%%%%%%%%%%%%%%%%%%%%%%%%%%%%%%%%%%%%%%%%%%
\begin{eqnarray}
\rho_\phi \propto a(t)^{- 3 (1 + w_\phi)}.
\label{rho_phi}
\end{eqnarray}
%%%%%%%%%%%%%%%%%%%%%%%%%%%%%%%%%%%%%%%%%%%%%%%%%%%%%%%%%%%%%%%%%%%

The Einstein equations are again given by Eq. (\ref{Einstein-eq 3}), and together with the present assumption
$\Lambda = 0$ and Eq. (\ref{rho_phi}), they are simply solved to 
%**   scale factor  %%%%%%%%%%%%%%%%%%%%%%%%%%%%%%%%%%%%%%%%%%%%%%%%%%%%%%%%%
\begin{eqnarray}
a(t) \propto t^{ \frac{2}{3 (1 + w_\phi)}}, 
\quad H(t) \equiv \frac{\dot{a}(t)}{a(t)} = \frac{2}{3 (1 + w_\phi)} \frac{1}{t}.
\label{scale factor}
\end{eqnarray}
%%%%%%%%%%%%%%%%%%%%%%%%%%%%%%%%%%%%%%%%%%%%%%%%%%%%%%%%%%%%%%%%%%%
Note that as promised, in this solution, the Hubble parameter is dependent on the time
variable $t$.

The remaining equations to be solved are the constraint equations (\ref{IG-constraints 1}) and field
equation (\ref{phi-eq 1}) for the scalar field $\phi$. For the last equation, the behavior of
the scalar field $\phi$ depends on the form of the potential $V(\phi)$, and since we do not specify a
concrete form of the potential, we proceed by simply assuming the existence of a scalar field 
which satisfies the field equation (\ref{phi-eq 1}).  Note that this assumption for the scalar field
is usually adopted in cosmology whenever we consider the equation of state $w_\phi = \frac{p_\phi}{\rho_\phi}$ 
with $w_\phi$ being a constant.

An important step for checking the presence of a classical solution is to find the solution satisfying
the equations (\ref{IG-constraints 1}). We will present the procedure of solving the equation,
$R = - \nabla_\mu \tau^\mu$ in detail below. 

As in Eq. (\ref{Spacial vector}), we can put $\tau_i$ to be zero by the gauge symmetries 
(\ref{Curved gauge transf}), and regard $\tau_0$ as an independent function which depends on only 
the time variable $t$, that is, $\tau_0 = \tau_0 (t)$. As seen in Eq. (\ref{Einstein-eq 3}), the Einstein
equations together with the conservation law give rise to the form of the Hubble parameter and its time derivative
%**   H & dot H  %%%%%%%%%%%%%%%%%%%%%%%%%%%%%%%%%%%%%%%%%%%%%%%%%%%%%%%%%
\begin{eqnarray}
H^2 = \frac{8 \pi G}{3} \rho_\phi, \quad \dot{H} = - 4 \pi G \left(1 + w_\phi \right) \rho_\phi.
\label{H & dot H}
\end{eqnarray}
%%%%%%%%%%%%%%%%%%%%%%%%%%%%%%%%%%%%%%%%%%%%%%%%%%%%%%%%%%%%%%%%%%%
Since we have $R = 6 ( 2 H^2 + \dot{H} )$ and $\nabla_\mu \tau^\mu = - \dot{\tau}_0 
- 3 H \tau_0$ in the metric ansatz (\ref{FRW metric}), the field equation $R = - \nabla_\mu \tau^\mu$
can be rewritten as
%**   R = d tau  %%%%%%%%%%%%%%%%%%%%%%%%%%%%%%%%%%%%%%%%%%%%%%%%%%%%%%%%%
\begin{eqnarray}
\dot{\tau}_0 + 3 H \tau_0 = 8 \pi G \left(1 - 3 w_\phi \right) \rho_\phi.
\label{R = d tau}
\end{eqnarray}
%%%%%%%%%%%%%%%%%%%%%%%%%%%%%%%%%%%%%%%%%%%%%%%%%%%%%%%%%%%%%%%%%%%
The time dependence of $\tau_0(t)$ can be expressed in terms of only the Hubble parameter $H(t)$,
so we can put 
%**   tau_0  %%%%%%%%%%%%%%%%%%%%%%%%%%%%%%%%%%%%%%%%%%%%%%%%%%%%%%%%%
\begin{eqnarray}
\tau_0 (t) = A H(t)^\alpha,
\label{tau_0}
\end{eqnarray}
%%%%%%%%%%%%%%%%%%%%%%%%%%%%%%%%%%%%%%%%%%%%%%%%%%%%%%%%%%%%%%%%%%%
where both $A$ and $\alpha$ are constants. Calculating the LHS of Eq. (\ref{R = d tau}) with the help
of (\ref{H & dot H}) and (\ref{tau_0}), we find that 
%**   R = d tau 2 %%%%%%%%%%%%%%%%%%%%%%%%%%%%%%%%%%%%%%%%%%%%%%%%%%%%%%%%%
\begin{eqnarray}
\dot{\tau}_0 + 3 H \tau_0 = 4 \pi G A \left( \frac{8 \pi G}{3} \right)^{\frac{1}{2} ( \alpha - 1)}
\left[ 2 - \alpha ( 1 + w_\phi) \right] \rho_\phi^{\frac{1}{2} ( \alpha + 1)}.
\label{R = d tau 2}
\end{eqnarray}
%%%%%%%%%%%%%%%%%%%%%%%%%%%%%%%%%%%%%%%%%%%%%%%%%%%%%%%%%%%%%%%%%%%
Comparing Eq. (\ref{R = d tau 2}) with Eq. (\ref{R = d tau}), we find that $\alpha = 1$ and 
$A = \frac{2 (3 w_\phi - 1)}{w_\phi -1}$. Hence, we reach the solution 
%**   tau_0 2 %%%%%%%%%%%%%%%%%%%%%%%%%%%%%%%%%%%%%%%%%%%%%%%%%%%%%%%%%
\begin{eqnarray}
\tau_0 (t) =  \frac{2 (3 w_\phi - 1)}{w_\phi - 1} H(t).
\label{tau_0 2}
\end{eqnarray}
%%%%%%%%%%%%%%%%%%%%%%%%%%%%%%%%%%%%%%%%%%%%%%%%%%%%%%%%%%%%%%%%%%%
In a perfectly similar way, one can find the solution for $\omega_0, \rho_0$ 
%**   omega_0, rho_0 %%%%%%%%%%%%%%%%%%%%%%%%%%%%%%%%%%%%%%%%%%%%%%%%%%%%%%%%%
\begin{eqnarray}
\omega_0 (t) =  \frac{2}{3 (w_\phi + 3)} \frac{1}{H(t)},
\quad \rho_0 (t) = - \frac{1}{4 \pi G} \frac{w_\phi}{w_\phi - 1} H(t).
\label{omega_0, rho_0}
\end{eqnarray}
%%%%%%%%%%%%%%%%%%%%%%%%%%%%%%%%%%%%%%%%%%%%%%%%%%%%%%%%%%%%%%%%%%%
Finally, using the explicit expression of the Hubble parameter in (\ref{scale factor}),
the solution reads
%**   Solution %%%%%%%%%%%%%%%%%%%%%%%%%%%%%%%%%%%%%%%%%%%%%%%%%%%%%%%%%
\begin{eqnarray}
\tau_0 (t) &=&  \frac{4 (3 w_\phi - 1)}{3 (w_\phi - 1)(w_\phi + 1)} \frac{1}{t},
\quad \omega_0 (t) =  \frac{w_\phi + 1}{w_\phi + 3} t,  \nonumber\\
\rho_0 (t) &=& - \frac{1}{6 \pi G} \frac{w_\phi}{(w_\phi - 1)(w_\phi + 1)} \frac{1}{t}.
\label{Solution}
\end{eqnarray}
%%%%%%%%%%%%%%%%%%%%%%%%%%%%%%%%%%%%%%%%%%%%%%%%%%%%%%%%%%%%%%%%%%%

Two remarks about our solution are in order. First, for $w_\phi = -1$, the conservation law
(\ref{Conservation-T2}) shows that $\rho_\phi$ is a constant, thereby the Hubble parameter becoming
a constant given by $H = \bar H = \sqrt{\frac{8 \pi G}{3} \rho_\phi}$.  Then the scale factor
has the exponential behavior $a(t) \propto \e^{\bar H t}$, which means the inflation universe.
The second remark is that for $- 1 < w_\phi < - \frac{1}{3}$ the scale factor in (\ref{scale factor})
shows that the universe is an acceleratingly expanding one like the present universe. It is of interest to
see that in this range of $w_\phi$, the solution (\ref{Solution}) is non-singular.

%%%%%%%%%%%%%%%%%%%%%%%%%%%%%%%%%%%%%%%%%%%%%%%%%%%%%%%%%%%%%%%%%%%%%
%%%%%%%%%%%%%%%%%%%%%%%%%%%%%%   SEC  5    %%%%%%%%%%%%%%%%%%%%%%%%%%
%%%%%%%%%%%%%%%%%%%%%%%%%%%%%%%%%%%%%%%%%%%%%%%%%%%%%%%%%%%%%%%%%%%%%
\section{Dynamical mechanism for fixing Lagrange multiplier fields}

Thus far, we have selected specific values (\ref{tau-omega-rho 3}) among arbitrary constants by hand,
which is obviously an unwelcome feature in the present approach. It is desirable that we can pick up
the values (\ref{tau-omega-rho 3}) through some dynamical mechanism.  

To clarify the issue to some degree, it is useful to take account of the partition function of 
the topological quantum field theory, which is given by
%**   Partition function %%%%%%%%%%%%%%%%%%%%%%%%%%%%%%%%%%%%%%%%%%%%%%%%%%%%%%%%%
\begin{eqnarray}
Z &=&  \int {\cal{D}} g_{\mu\nu} {\cal{D}} \phi {\cal{D}} \gamma {\cal{D}} \lambda
{\cal{D}} \beta {\cal{D}} \tau^\mu {\cal{D}} \omega^\mu {\cal{D}} \rho^\mu {\cal{D}} b^{(i)}
{\cal{D}} c^{(i)} \ \e^{i S_T}
\nonumber\\
&=& \int {\cal{D}} g_{\mu\nu} {\cal{D}} \phi {\cal{D}} \bar \gamma {\cal{D}} \bar \lambda
{\cal{D}} \bar \beta \left[ \det (\nabla_\mu \nabla^\mu) \right]^3 
\e^{i \int d^4 x \sqrt{-g} \left\{\bar \gamma R + \bar \lambda + \bar \beta 
\left[ -\frac{1}{2} g^{\mu\nu} \nabla_\mu \phi \nabla_\nu \phi - V(\phi) \right] \right\} },
\label{Partition function}
\end{eqnarray}
%%%%%%%%%%%%%%%%%%%%%%%%%%%%%%%%%%%%%%%%%%%%%%%%%%%%%%%%%%%%%%%%%%%
where $\bar \gamma, \bar \lambda, \bar \beta$ are constant modes. Though we have picked up the
specific values (\ref{tau-omega-rho 3}) for these constants,  as can be seen in the partition function 
(\ref{Partition function}), there are in principle no preferred values 
for such constants, so the real problem is how to pick up these physically plausible vaules 
through some dynamical mechanism in the framework of quantum field theories. Another annoying problem,
which is also manifest in the partition function (\ref{Partition function}), is the presence of 
the determinant factor coming from the ghosts. 

It might be premature to present one solution to overcome the above problems, but nevertheless 
we cannot help presenting it for future developments of the approach at hand. A toy model in a flat 
Minkowski space-time again provides us with a useful playground for proposing our idea
%**   New toy model 1   %%%%%%%%%%%%%%%%%%%%%%%%%%%%%%%%%%%%%%%%%%%%%%%%%%%%%%%%%
\begin{eqnarray}
{\cal{L}} = \gamma \left[ \Psi(\varphi) + \partial_\mu \tau^\mu \right] 
+ \frac{1}{2} \tau_\mu \tau^\mu - \frac{f}{4} \left( \gamma^2 - \bar \gamma^2 \right)^2,
\label{New toy model 1}
\end{eqnarray}
%%%%%%%%%%%%%%%%%%%%%%%%%%%%%%%%%%%%%%%%%%%%%%%%%%%%%%%%%%%%%%%%%%%
where $f$ is a coupling constant. Note that in order to avoid the problem of the topological ghost determinant,
we have worked with not a topological field theory but a pure bosonic theory. Also note that
we have added the quadratic term of the vector field $\tau^\mu$ and the Higgs potential 
for the scalar field $\gamma$ to trigger the spontaneous symmetry breakdown. 

Now, integrating over the vector field $\tau^\mu$ exactly gives us an effective theory 
%**   New toy model 2   %%%%%%%%%%%%%%%%%%%%%%%%%%%%%%%%%%%%%%%%%%%%%%%%%%%%%%%%%
\begin{eqnarray}
{\cal{L}} = \gamma \Psi(\varphi) - \frac{1}{2} \partial_\mu \gamma \partial^\mu \gamma
- \frac{f}{4} \left( \gamma^2 - \bar \gamma^2 \right)^2,
\label{New toy model 2}
\end{eqnarray}
%%%%%%%%%%%%%%%%%%%%%%%%%%%%%%%%%%%%%%%%%%%%%%%%%%%%%%%%%%%%%%%%%%%
which is the standard Lagrangian density of the scalar field with the Higgs potential 
except the first term. Thus, at the leading order, $\langle \gamma \rangle = \bar \gamma$
where $\langle \gamma \rangle$ means the vacuum expectation value, the Lagrangian
density (\ref{New toy model 2}) simply takes the form
%**   New toy model 3   %%%%%%%%%%%%%%%%%%%%%%%%%%%%%%%%%%%%%%%%%%%%%%%%%%%%%%%%%
\begin{eqnarray}
{\cal{L}} = \bar \gamma \Psi(\varphi).
\label{New toy model 3}
\end{eqnarray}
%%%%%%%%%%%%%%%%%%%%%%%%%%%%%%%%%%%%%%%%%%%%%%%%%%%%%%%%%%%%%%%%%%%
In this way, we can reach a theory with a fixed value $\bar \gamma$ for the scalar field $\gamma$
via the spontaneous symmetry breakdown.

Armed with this toy model, we are now ready to present the full Lagrangian density for the
present theory
%**   New action 1  %%%%%%%%%%%%%%%%%%%%%%%%%%%%%%%%%%%%%%%%%%%%%%%%%%%%%%%%%
\begin{eqnarray}
\frac{1}{\sqrt{-g}} {\cal{L}} &=& \gamma \left( R + \nabla_\mu \tau^\mu \right)
+ \frac{1}{2} \tau_\mu \tau^\mu - \frac{f_\gamma}{4} \left( \gamma^2 - \bar \gamma^2 \right)^2
\nonumber\\
&+& \lambda \left( 1 + \nabla_\mu \omega^\mu \right)
+ \frac{1}{2} \omega_\mu \omega^\mu - \frac{f_\lambda}{4} \left( \lambda^2 - \bar \lambda^2 \right)^2
\nonumber\\
&+& \beta \left[ - \frac{1}{2} g^{\mu\nu} \nabla_\mu \phi \nabla_\nu \phi 
- V(\phi) + \nabla_\mu \rho^\mu \right]
+ \frac{1}{2} \rho_\mu \rho^\mu - \frac{f_\rho}{4} \left( \beta^2 - \bar \beta^2 \right)^2,
\label{New action 1}
\end{eqnarray}
%%%%%%%%%%%%%%%%%%%%%%%%%%%%%%%%%%%%%%%%%%%%%%%%%%%%%%%%%%%%%%%%%%%
where $f_\gamma, f_\lambda, f_\rho$ are coupling constants, and $\bar \gamma, \bar \lambda, \bar \beta$
are defined in (\ref{tau-omega-rho 3}). As in the toy model, integrating over the vector fields 
$\tau^\mu, \omega^\mu, \rho^\mu$ and taking the leading values of the vacuum expectation values
$\langle \gamma \rangle = \bar \gamma, \ \langle \lambda \rangle = \bar \lambda, \ \langle \beta \rangle 
= \bar \beta$, we can reproduce the desired Lagrangian density
%**   New action 2  %%%%%%%%%%%%%%%%%%%%%%%%%%%%%%%%%%%%%%%%%%%%%%%%%%%%%%%%%
\begin{eqnarray}
\frac{1}{\sqrt{-g}} {\cal{L}} &=& \bar \gamma R + \bar \lambda 
+ \bar \beta \left[ - \frac{1}{2} g^{\mu\nu} \nabla_\mu \phi \nabla_\nu \phi - V(\phi) \right]
\nonumber\\
&=& \frac{1}{16 \pi G} \left( R - 2 \Lambda \right)
- \frac{1}{2} g^{\mu\nu} \nabla_\mu \phi \nabla_\nu \phi - V(\phi),
\label{New action 2}
\end{eqnarray}
%%%%%%%%%%%%%%%%%%%%%%%%%%%%%%%%%%%%%%%%%%%%%%%%%%%%%%%%%%%%%%%%%%%
where at the last expression, we have used (\ref{tau-omega-rho 3}).

%%%%%%%%%%%%%%%%%%%%%%%%%%%%%%%%%%%%%%%%%%%%%%%%%%%%%%%%%%%%%%%%%%%%%
%%%%%%%%%%%%%%%%%%%%%%%%%%%%%%   SEC  6    %%%%%%%%%%%%%%%%%%%%%%%%%%
%%%%%%%%%%%%%%%%%%%%%%%%%%%%%%%%%%%%%%%%%%%%%%%%%%%%%%%%%%%%%%%%%%%%%
\section{Conclusion}

In this article, motivated with unimodular gravity \cite{Einstein}-\cite{Carballo-Rubio}, 
we have developed a new type of induced gravity.
In unimodular gravity \cite{Einstein}-\cite{Carballo-Rubio}, the cosmological constant emerges 
as an integration constant which is unrelated to parameters existing in the original action
whereas in our induced gravity, the Newton constant, the cosmological constant and the coefficient 
in front of the matter action appear as integration constants. This physically intriguing phenomenon 
is utilized to generate general relativity with matter field by starting with the induced gravity 
and solving its field equations at the classical level. 
In our approach, we have never appealed to quantum fluctuations of matter (or gauge field) to create
the dynamics of general relativity and matter.
We therefore think that our induced gravity is a natural counterpart in a local field theory
for the pregeometrical approach of string field theory.

Moreover, we have advocated our original idea such that the induced gravity made in this article 
is closely connected with a topological quantum field theory. To do so, we have analysed the constraint
system of the induced gravity in the canonical formalism in detail, and shown that in order to nullify 
a ghost field existing in this theory, a topological symmetry plays a critical role and the induced gravity 
must be deformed into a topological quantum field theory. It is worthwhile to stress that the topological 
symmetry is needed to make a physically consistent theory in a natural way.   

We have also examined classical solutions, which satisfy all the field equations of the topological
quantum field theory obtained from the induced gravity. It is clarified that not only the inflation universe
but also the acceleratingly expanding universe can emerge as classical solutions. 

In this article, we have treated with the metric tensor field as a fixed classical background
field along the spirit of the induced gravity, but it might be interesting to regard
the metric tensor as a quantized field as well and study its physical implications. 
\footnote{We have already made models of topological gravity in four dimensions 
\cite{Oda5}-\cite{Oda7}.}

There are many of problems to be clarified in future. One of the most important problems is
to undestand the quantum effects of the present theory. In particular, we have confined ourselves 
to only the leading-order treatment of spontaneous symmetry breaking in Section 5, so we must
take the quantum fluctuations around the vacuum expectation values into consideration. 
Another more difficult problem is of course to study the well-known cosmological constant
problem within the present formalism. We wish to return these problems in near future.

%%%%%%%%%%%%%%%%%%%%%%%%%%%%%%%%%%%%%%%%%%%%%%%%%%%%%%%%%%%%%%%%%%
%%%%%%%%%%%%%%%%%%%%%%%% Acknowledgements %%%%%%%%%%%%%%%%%%%%%%%%%%%%%
%%%%%%%%%%%%%%%%%%%%%%%%%%%%%%%%%%%%%%%%%%%%%%%%%%%%%%%%%%%%%%%%%%
\begin{flushleft}
{\bf Acknowledgements}
\end{flushleft}
This work is supported in part by the Grant-in-Aid for Scientific 
Research (C) No. 25400262 from the Japan Ministry of Education, Culture, 
Sports, Science and Technology.

%%%%%%%%%%%%%%%%%%%%%%% reference %%%%%%%%%%%%%%%%%%%%%%%%%%%%%%%
%%%%%%%%%%%%%%%%%%%%%%%%%%%%%%%%%%%%%%%%%%%%%%%%%%%%%%%%%%%%%%%%%%

\end{document}